\documentstyle[preprint,aps]{revtex}

\begin{document}
\draft
\preprint{UM-ChE-96/004}
\title{Universality of finite-size corrections to the number of critical
percolation clusters}
\author{Robert M. Ziff,$^1$  Steven R. Finch,$^2$ and Victor S. Adamchik$^3$}
\address{
$^1$ University of Michigan, Ann Arbor, MI 48109-2136}
\address{$^2$ MathSoft, Inc., Cambridge, MA  02142-1521}
\address{$^3$ Wolfram Research, Inc.,  Champaign, IL  61820-7237}

\date{\today}
\maketitle
\begin{abstract}

Monte-Carlo simulations on a variety of 2d percolating
systems at criticality suggest that the excess number of clusters in finite systems
over the bulk value $n_c$ is a universal quantity,
dependent upon the system shape but
independent of the lattice and percolation type.
Values of $n_c$ are found to high
accuracy, and for bond percolation confirm the theoretical predictions
of Temperley and Lieb, and Baxter, Temperley and Ashley,
which we have evaluated explicitly in terms
of simple algebraic numbers.
Predictions for the fluctuations are also verified for the first time.
\end{abstract}

\narrowtext
The standard percolation model \cite{SA} involves the random
occupation of sites or bonds of a regular lattice.
At a critical occupation probability $p_c$,
the mean size of clusters of adjacently occupied
sites becomes infinite, signalling the percolation
transition.  
The number $n(p)$ of clusters per site, however,  remains
finite, and attains a value $n_c = n(p_c)$
with critical behavior
$|p-p_c|^{2-\alpha}$ as $p \to p_c$, where
$2-\alpha = 8/3$.  Critical exponents
like $\alpha$ are universal,
having the same value for all systems of a given
dimensionality no matter what the lattice or percolation type,
essentially because they derive from
the large-scale, fractal properties of the percolation 
clusters.

On the other hand, $n_c$, like $p_c$, is a non-universal
quantity which varies from system to system.  This is because 
it depends upon the microscopic, lattice-level
behavior of the systems.  For some systems, $n_c$ is known theoretically
\cite{TL,BTA},
while for others, it must be determined 
by Monte-Carlo (MC) means.  Accurate values of $n_c$ --- a seemingly
basic property of a percolating system --- do not seem
to be available for many common lattices.  Thus we embarked
on a project to determine it for a variety of systems
by MC simulation.

To obtain accurate values, we employed a very
extensive set of simulations.
They  were carried out on square (SQ)
and triangular (TR) lattices,
with both site (S) and  bond (B) percolation, and the honeycomb (HC) lattice, for
B only.  Various orientations and system
boundaries were used as shown in Fig.~1. 
Periodic boundary conditions were used.
Systems ranged from $16\times16$ to
$512\times512$ in size, and between 
$10^6$ to $10^9$ samples were generated for each, which required
a total of several months of workstation computer time.
All but one set of simulations were carried out
at the critical point $p_c$ of the respective lattice. 

Because simulations must necessarily be done on finite systems, an
essential part of such a project is to determine the
nature of the finite-size corrections.
The use of periodic boundary conditions reduces
those corrections to sub-surface terms.
In characterizing the remaining corrections, we found a
surprising result: for systems of a given shape, the finite-size
corrections to $n_c$ appear to follow a universal behavior.
For a finite system of $S$ sites, we found that the average
density of clusters (number per site) behaves as
\begin{equation}
n = n_c + { b \over S} + \ldots
\label{nS}
\end{equation}
where $b$ is a function of the system shape only
and thus appears to be a universal quantity. 
For example, the results for a system with a square
boundary of dimension $L \times L$
are shown in Fig.~2, where we plot $n(S)$ vs.\ $1/S$.
  For this boundary shape
we considered three systems: S-SQ-I (site percolation
on a square lattice with orientation I shown in 
Fig.~1), B-SQ-I, and B-SQ-II, where II is a
square lattice rotated by 45$^\circ$.
The equality of the slopes of the three curves
means that $b$ has the same
value $\approx 0.8835$ for all these systems.
Higher-order corrections to (\ref{nS}) are
small and not visible above the statistical errors
of our simulations.

Considering rectangular systems of dimension $L \times L'$,
 we found that
$b$ increases monotonically with increasing aspect ratio
$r = L'/L$ as shown in Table I. \ 
Here, we considered only
one type of system for each aspect ratio.
Besides B-SQ, we also considered B-TR
with the boundary III.  With III's rectangular boundary,
a normal torus is produced when the periodic
boundary conditions are applied, and the effective aspect
ratio of the system we used is $\sqrt{3}$.
Note that, because $b(r) = b(1/r)$, the value $b(1) \approx 0.8835$
is a minimum for rectangular systems.

Besides rectangular systems, we also considered a
60$^\circ$ rhomboid system with an effective helical or ``twisted"
periodic boundary conditions (systems IV and V).
This type of boundary results when a triangular lattice
is represented in the computer
by a square lattice with one set of diagonals drawn in.
The periodic boundary conditions are applied to the
lattice in the squared-off orientation, and an effective helical
twist is introduced in the torus when the lattice
is transformed back to its true configuration
with the triangles equilateral.
For such a system with sides $L \times L$, we find $b \approx 0.878$,
somewhat smaller than the minimum value above for a rectangular boundaries.
(Likewise, 0.878 should be the minimum value of $b$ for all 60$^\circ$
parallelograms.)
Note that the effective height-to-width of this
system is $\sqrt{3}/2$, although it is not really comparable
to a rectangular system of a given aspect ratio because of the twist
in the boundary condition.

While we do not have a proof that $b$ is universal, we can make
the following heuristic arguments to support that hypothesis.
Eq.~(\ref{nS}) can be written
\begin{equation}
N \approx n_c S + b 
\label{NS}
\end{equation}
where $N = n S $ is the total number of clusters in the system.
This equation shows that $b$ is effectively the excess number of clusters
over the expected bulk value, $n_c S$.  As such, it represents large clusters
whose size is of the order of the size of the system.  While the number
and distribution of small, ``microscopic" clusters is non-universal,
the distribution of these larger clusters is.
One can also relate $b$ to the average number
$w$ of clusters wrapping around the toroidal system, which too is 
a universal quantity.
Consider taking one $2L \times 2L$ system,
unwrapping the periodic boundary conditions,
cutting it into four $L \times L$ systems, and then re-applying
the periodic boundary conditions to each of these.
The number of clusters $N(L)$ should roughly satisfy
$N(2L) - w = 4(N(L)-w)$, since the density of non-wrapping
clusters will be about the same, and this formula
is consistent with (\ref{NS}) with $w= b$ (here, $S = L^2$).
A similar argument shows that
$b$ should increase as the aspect ratio $r = L'/L$ increases,
since there are then more wraparound clusters, consistent
with what we have observed.

The values of $n_c$ that we found from plots like Fig.~2
are listed in Table II.   The results for the bond percolation
systems are consistent with theoretical predictions that have been
made for these systems.
For B-SQ, where $p_c = 1/2$, Temperley and Lieb \cite{TL} showed 
\begin{equation}
n_c^{B\hbox{-}SQ} = {1\over2}\left[Z  {\partial I_1
\over \partial Z}\right]_{Z=1}
= {1\over2}\left[- \cot \mu {\partial I_1 \over \partial \mu}\right]_{\mu = \pi/3}
\label{TL1}
\end{equation}
where $Z = 2 \cos \mu$ and 
\begin{equation}
I_1 =
{1 \over 4\mu}\int_{-\infty}^{\infty}{\rm sech}
\Big({\pi\alpha\over2\mu}\Big)\ln\Big({\cosh\alpha-\cos2\mu\over\cosh\alpha-1}\Big)
{\rm d}\alpha \ ,
\label{TL2}\end{equation}
and reported the numerical estimate $0.0980_7$.
In fact, we have found that this rather complicated integral result
can be evaluated explicitly \cite{victorsq}.
First, integrate (\ref{TL2}) by parts to obtain
\begin{equation}
I_1 =
{4 \sin^2 \mu \over \pi}\int_{0}^{\infty}
{\tan^{-1} \big(\tanh {\pi\alpha\over4\mu} \big) \coth {\alpha\over 2}
\over
\cosh \alpha  - \cos 2 \alpha  } d\alpha 
\label{TL3}
\end{equation}
and then differentiate to find:
\begin{eqnarray}
{\partial I_1 \over \partial \mu}\Bigg |_{\mu= \pi/3}
&&= {8 \sqrt{3} \over \pi} 
\int_{0}^{\infty}
{\tan^{-1}
\big(\tanh {3\alpha\over4} \Big) \sinh \alpha \over
(2 \cosh \alpha + 1)^2 }d\alpha \nonumber \\
&&-{27\over \pi^2} \int_{0}^{\infty}
{\alpha \cosh {\alpha\over2} \, {\rm csch}\,3 \alpha }\,d \alpha
\label{TL4} \ .
\end{eqnarray}
Substituting 
$\alpha = - 2 \log z$ in the first integral above, we can reduce it
to a form that can be evaluated with the help of Mathematica
\cite{math},
yielding $2 \sqrt3 - 3$.
Likewise, the second integral in (\ref{TL4}) can also be evaluated,
and leads to the simple final result
\begin{equation}
n_c^{B\hbox{-}SQ}={3\sqrt{3}-5\over2} \approx 0.098\,076\,211 \ .
\label{V1}
\end{equation}
Further details are given in \cite{victorsq}.
Our numerical results in Table II are consistent with this prediction
to all significant figures. 
Likewise, for B-TR,
where $p_c^{B\hbox{-}TR} = 2 \sin (\pi/18)$ is the solution
to $p^3 - 3p + 1 = 0$ \cite{SE64}, 
Baxter, Temperley and Ashley \cite{BTA} showed that
\begin{equation}
n_c^{B{\rm -}TR} = \left[- {\csc 2\phi \over 4}
{\partial I_2 \over \partial \phi}\right]_{\phi = \pi/3} 
+ {3 \over 2} - {2 \over 1+p_c^{B\hbox{-}TR}}  
\label{BTA1}
\end{equation}
where
\begin{equation}
I_2 =
 {3 \over 2}\int_{-\infty}^{\infty}
 {\sinh(\pi-\phi)x\, \sinh {2\over 3}\phi x \over x \sinh\pi x \, \cosh\phi x}dx \ ,
 \label{BTA2}
\end{equation}
and estimated $n_c \approx 0.1118$ by numerical integration and also
from the 16-term series available at that time.
Again, we have explicitly evaluated this expression \cite{victortri,victormaui},
with the result 
\begin{equation}
n_c^{B\hbox{-}TR} = {35 \over 4} - {3 \over p_c^{B\hbox{-}TR}} \approx 0.111\,844\,275
\end{equation}
which is in agreement with our observations. 

Making use of duality, we also studied the B-HC system
at the same time as B-TR.
Our value of $n_c$ for this lattice given in Table II
agrees with prediction
$n_c^{B\hbox{-}HC} = n_c^{B\hbox{-}TR} + (p_c^{B\hbox{-}TR})^3 \approx 0.153\,733\,341$,
which
follows from the results of \cite{ES66} written in
terms of clusters of wetted sites
per site on the TR lattice.
Note that for all bond
percolation systems, we count the number of clusters of all wetted sites, 
which includes both clusters of connected sites and ``null" clusters
of isolated sites with no occupied bonds attached to them.  
Having the above exact results for
$n_c$ was very useful for finding $b$ to high accuracy for these systems.

Previous numerical results are also consistent with our findings.
For B-SQ, Nakanishi and Stanley \cite{NS} found 0.98075 by simulation.
Domb and Pearce \cite{DP} found $n_c = 0.0173(3)$ by series analysis, where
here $n_c$ is 
the number of bond clusters per bond.  To put (\ref{V1}) in this form,
one must subtract the concentration of isolated sites (1/16)
and then divide by two, the number of bonds per site.  This yields

\begin{equation}
n_c^{B\hbox{-}SQ}({\hbox{per bond}}) = {24 \sqrt{3} - 41\over 32} \approx 0.017\,788\,106,
\label{bond-bond}
\end{equation}
and shows that the results of \cite{DP} are slightly low.
Surprisingly, we have found little comparison with or
discussion of the results of
\cite{TL} and \cite{BTA}
in the percolation literature --- a notable exception being \cite{Essam},
where in fact we first became aware of \cite{TL}.

For S-SQ, where $p_c = 0.592\,746\,0(5)$ \cite{ziff92},
we find
\begin{equation}
n_c^{S\hbox{-}SQ}=0.027\,5931\pm 0.000\,0003 \ ,
\label{SSQ} 
\end{equation}
but could not
find any published values to compare this with.
A simple Pad\'e analysis \cite{Adler} of 
the 25-th order series for $n(p)$ given in \cite{CG},
with no attempt made to account for the branch-point singularity at $p_c$,
yields $n_c =0.027\,54(4)$, the error representing the variation 
for [N,D] = [12,13], [13,12],$\ldots$.

For S-TR lattice, where $p_c = 1/2$, we find
\begin{equation}
n_c^{S\hbox{-}TR} = 0.017\,6255\pm 0.000\,0005  \ .   
\label{STR} 
\end{equation}
This result is consistent with the
previous MC value $0.017\,630 (2)$
of Margolina et al. \cite{MNSS} but not the early low-order
series value $0.0168(2)$ \cite{DP}.
As in the S-SQ case, no theoretical prediction for this quantity exists.
Note that the value of $n_c^{S\hbox{-}TR}$ is quite close
to $n_c^{B\hbox{-}SQ}$ of (\ref{bond-bond}),
which is reasonable because
the matching site lattice to the B-SQ system is quite similar to the S-TR system,
both having coordination number 6 and $p_c = 1/2$.

We also considered S-SQ at $p = 0.5$,
well below $p_c$.
Our MC results give
\begin{equation}
n^{S\hbox{-}SQ}(0.5) = 0.065\,770\,3 \pm 0.000\,000\,2 \ .   
\label{SQB5} 
\end{equation}
This quantity relates to a simple question of graph theory \cite{finch}: how
many clusters of 1's exist on a matrix filled randomly with 0's and 1's of equal
concentration?  Here finite-size effects were observed in lattices
from $32\times32$ to $512\times512$ in size (i.e., $b = 0$),
presumably because the
dimension of a typical cluster was smaller than 32.  
In this case, a straightforward Pad\'e analysis of Conway and Guttmann's series \cite{CG}
yields a result
of nearly the same accuracy:
$0.065\,769\,6 (6)$.

The fluctuations $\sigma^2$ in the number
of clusters from sample to sample,
determined by 
$\langle (\Delta N)^2 \rangle =
\langle N^2 \rangle - \langle N \rangle^2 = S\sigma^2$ for large $S$,
are given in Table II.  \ 
Like $n_c$, 
$\sigma^2$ is a function of the lattice and percolation type
and thus non-universal, but still independent
of the boundary condition and shape.   The value 0.164 found for B-SQ, for example, was 
found for square systems with both orientations, rectangular systems of aspect
ratio 2 and 4, and also a square system with open boundaries. 
The error (one standard deviation) in 
the measured value of $n$ is $\sigma(S N_{runs})^{-1/2}$
where $N_{runs}$ is the number of samples; this is how
we obtained the error bars in our numerical results above.
Note that previously, only rough measurements of fluctuations were made \cite{CSS}.

For the B-SQ case, the theory of \cite{TL} yields a prediction for the fluctuations
of a quantity $H \equiv C_G + S_G$, where $C_G$ is the number of components
(clusters plus isolated sites --- the same as $N$ above)
and $S_G$ is the number of independent cycles in the 
system:
\begin{equation}
{\langle(\Delta H)^2\rangle \over S}
= \left[\left(\cot \mu {\partial  \over \partial \mu}\right)^2 I_1\right]_{\mu = \pi/3}
\label{fluc1}
\end{equation}
The  numerical results given in \cite{TL} (p.\,280) imply the value
$0.196_2 - 0.037_7 = 0.158_5$ for this quantity.
After many intermediate steps, we have evaluated the resulting integral
expressions explicitly and find \cite{victorsq}
\begin{equation}
{\langle(\Delta H)^2\rangle \over S} = {8\sqrt{3} - 25 \over 2} + {18\over\pi}
\approx 0.157\, 781\, 182 \ .
\label{fluc2}
\end{equation}
At $p_c$, $S_G = C_G$, but because
of the cross term $\langle C_G S_G \rangle$ in
$\langle(\Delta H)^2\rangle$, the above result
cannot be used to obtain a prediction for $\sigma^2 = \langle (\Delta C_G)^2\rangle/S$. 
However, one can show that $S_G$ is equal
to the number of components on the dual lattice $C_{\tilde G}$, which 
can be measured at the same time as clusters on $C_{G}$.
  (It also follows that 
$H$ is equal to the total number of hulls that can be drawn on the system,
minus 1 if there is wraparound in both directions \cite{pinson}.) 
Indeed, for B-SQ percolation on rectangular systems, we find that
$\langle(\Delta H)^2\rangle = 0.1578(1)$, in perfect agreement
with (\ref{fluc1}).  
If $C_G$ and $C_{\tilde G}$ were uncorrelated,
then ${\langle(\Delta H)^2\rangle / S}$
would equal $2\sigma^2=0.328$.  This implies that
realizations with more
clusters on the regular lattice tend to have fewer
clusters on the dual lattice, and
{\it vice versa}, as one might expect.  
The fluctuations of $H$ have finite-size corrections that behave
as $\approx L^{-2}$, while the fluctuations $\sigma^2$
of clusters alone  was found to follow an unusual behavior
 $\approx L^{-1.25}$.

We have also measured $ \langle(\Delta H)^2\rangle / S $ for
the B-TR case, where now $C_{\tilde G}$
are components on the B-HC lattice,
and find $0.2207(1)$.
We have not related this result to theory \cite{BTA};
although we can evaluate the next derivative of $I_2$ \cite{victortri}, we
have not found the additional terms analogous to the last two terms in (\ref{BTA1}).

Thus, we have shown that the theory of \cite{TL} can be applied to 
cluster fluctuations if one includes
the dual-lattice clusters in the measurements.
We have found surprisingly simple expressions for the integrals of the
theories of \cite{TL,BTA}, which we have also verified numerically to
high precision.
And we have presented evidence that the excess number of clusters is a universal
quantity, much like the crossing probability that has been of much interest
recently \cite{pinson,Cardy,Langlands,Aizenman,Watts}.
Recently, Aharony and Stauffer \cite{AS} have presented theoretical arguments
for the universality of $b$, and furthermore, Kleban \cite{Kleban} has
derived explicit expressions for $b$ using conformal
invariance methods, which support this universality hypothesis.

The authors thank P. Kleban, D. Stauffer and A. Aharony for their comments.
This material is based in part upon work supported by the US 
National Science Foundation under Grant No.\thinspace DMR-9520700.

\begin{figure}
\caption{ 
Different lattices and boundary shapes used in the MC studies.
Systems I, II and III have rectangular 
boundaries and lead to
normal tori when the periodic boundaries are applied.
IV and V have a rhomboid boundary,
but the periodic boundary conditions are
applied when the lattice is in a squared-off form,
leading to a helical or twisted boundary. }

\end{figure}

\begin{figure}
\caption{A plot of $n = N/S$ vs.
$ 1/S$ for the $L\times L$
S-SQ-I (filled square), B-SQ-I (open square) and
B-SQ-II (open diamond) systems at $p_c$.
The curves are offset vertically for clarity;
the actual intercepts are $n_c^{S{\hbox{-}}SQ}$ for the upper
curves, and  $n_c^{B{\hbox{-}}SQ}$ for the lower two curves.
This plot illustrates that $b$ (the slope) is the same
for all three systems.}
\end{figure}

\vfill\eject

\begin{table}
\caption{Measured values of the universal finite-size correction constant $b = N - n_c S$,
grouped by universality class (system shape and boundary condition).
Numbers in 
parentheses show the errors in the last digit(s). 
\label{table1}}
\begin{tabular}{lll}
system&boundary&$b$\\
\tableline
\tableline
S-SQ-I&square&$0.8832(3)$\\
B-SQ-I&square&$0.8838(5)$\\
B-SQ-II&square&$0.8835(8)$\\
\tableline
B-TR-III&$1\times\sqrt{3}$ rectangle&$0.946(2)$\\
\tableline
B-SQ-II&$1\times 2$ rectangle&$0.991(2)$\\
\tableline
B-SQ-II&$1\times 4$ rectangle&$1.512(3)$\\
\tableline
S-TR-IV&$1 \times 1$ rhombus&$0.8783(8)$\\
B-TR-IV&$1 \times 1$ rhombus&$0.878(1)$\\
B-HC-V&$1 \times 1$ rhombus&$0.877(1)$\\
\end{tabular}
\end{table}

\begin{table}
\caption{MC results for cluster concentration
$n_c$ and fluctuations $\sigma^2$ for the
various systems studied, grouped by system type, along
with known values of $p_c$ [6,12].  
These quantities are non-universal,
but independent of the system boundary, as
seen here for the B-SQ and B-TR cases.
\label{table2}}
\begin{tabular}{lllll}
system&boundary&$p_c$&$n_c$&$\sigma^2$\\
\tableline
\tableline
S-SQ-I&square&$0.592\,746$&$0.027\,5981(3)$&0.05385\\
\tableline
S-TR-IV&$1 \times 1$ rhombus &$0.5$&$0.017\,6255(5)$&0.0309\\
\tableline
B-SQ-I&square&$0.5$&$0.098\,0763(8)$&0.1644\\
B-SQ-II&square&$0.5$&$0.098\,0765(10)$&0.1644\\
B-SQ-II&$1\times 2$ rectangle&$0.5$&$0.098\,076$&0.1642\\
B-SQ-II&$1\times 4$ rectangle&$0.5$&$0.098\,1$&0.163\\
\tableline
B-TR-III&$1\times\sqrt{3}$ rect.&$0.347\,296$&$0.111\,846(2)$&0.183\\
B-TR-IV&$1 \times 1$ rhombus&$0.347\,296$&$0.111\,843(2)$&0.1827\\
\tableline
B-HC-V&$1 \times 1$ rhombus&$0.652\,705$&$0.153\,735(2)$&0.267\\
\end{tabular}
\end{table}

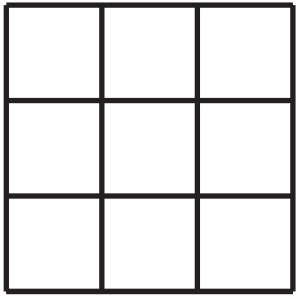
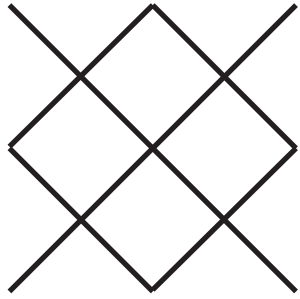
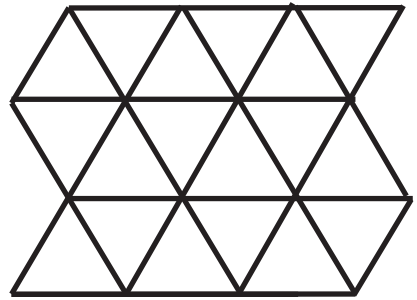
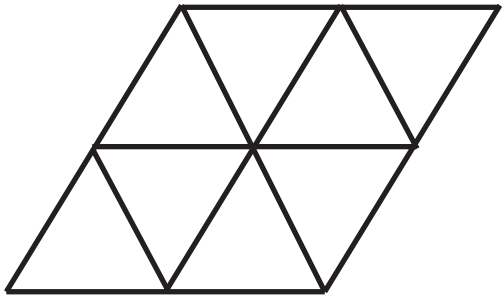
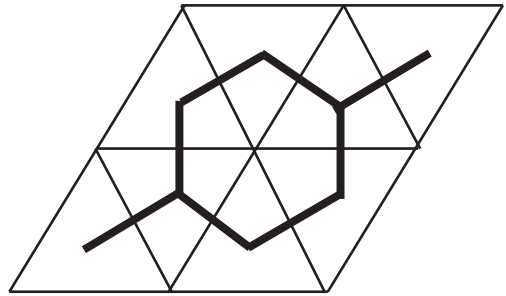
\end{document}